# New enlargement of the novel class of superconductors

V.L. Kozhevnikov, O.N. Leonidova, A.L. Ivanovskii, I.R Shein, B.N. Goshchitskii*, A.E. Karkin*

Institute for Solid State Chemistry, Ural Branch of the Russian Academy of Sciences, Yekaterinburg 620041, GSP-145, Russia
* Institute for Metals Physics, Ural Branch of the Russian Academy of Sciences, Yekaterinburg 620041, GSP-170, Russia

**Here we report on synthesis and characterization of a new member $LaO_{1-\delta}NiBi$ in the family of novel superconductors. Though the onset superconducting temperature ~4 K is smaller compared to 55 K that has been achieved already in $SmO_{1-\delta}FeAs$ the similarities in crystalline structure, electronic properties and low-temperature electron transport features give evidence to the common mechanism that seems to underlie the superconducting state in lanthanum nickel oxybismuthides and oxyarsenides.**

E-mail: kozhevnikov@ihim.uran.ru

New layered superconductor, fluorine doped iron oxyarsenide LaOFeAs with $T_C$ of about 26K, was reported in February 2008 [1]. This finding has attracted a great deal of interest because it seems to have set a first example of long sought for copper-free compounds with properties combination that suggest unconventional superconductivity [2-7]. Also it has triggered much activity in search of related materials. Doping and chemical substitutions are known to be powerful tools for properties tailoring, and it is not surprising that soon after these methods have resulted in discovery of strontium doped derivatives $La_{1-x}Sr_xOFeAs$ [8]. Further substitution of lanthanum for other rare earths (Re) has resulted in finding of fluorine doped $ReO_{1-x}F_xFeAs$ [9-11] and oxygen deficient $ReO_{1-\delta}FeAs$ [12] where the $T_c$ onset for Re = Sm was observed at 55 K. Reports have been published also on synthesis of LaOFeP [13] and LaNiOP [14]. The critical temperature in these compounds is about 4 K only. Still these finding are important because they demonstrate that the role played by arsenic is not unique and it can be replaced with other less dangerous elements.

The basic quaternary oxyarsenide LaOFeAs adopts a layered tetragonal crystal structure (ZrCuSiAs type, space group P4/*nmm*, Z = 2 [15]), where layers of (La-O) are sandwiched with (Fe-As) layers. According to DFT calculations [5,16-18] the electron bands in LaOFeAs near the Fermi level are formed mainly by the states of (Fe-As) layers while the bands of (La-O) layers are rather far from the Fermi level. Some experimental [19] and theoretical [20] attempts have been made already for understanding composition dependent behavior of oxyarsenides ReOFeAs and LaOMAs (M = V, Cr…Ni, Cu) [21].

It is shown that variations in inter- and intra-layer distances and in chemical composition all seem to be important factors for properties' optimization in these materials. In this communication we report on synthesis and characterization of $LaO_{1-\delta}NiBi$.

As a preliminary part we carried out band structure calculations for LaOMBi, where M = Sc, Ti…Ni, Cu, with the help of the full potential method with mixed basis APW+lo (LAPW) implemented in the WIEN2k suite of programs [22]. The generalized gradient correction (GGA) to exchange-correlation potential in PBE form [23] was used. The calculations were performed with full lattice optimizations including internal coordinates, $z_{La}$ and $z_{Bi}$, which define inter-layer distances La-O and M-Bi, respectively [15]. The self-consistent calculations were considered to have converged when difference in the total energy of the crystal did not exceed 0.01 mRy as calculated at consecutive steps. The calculations were performed for nonmagnetic (NM) and ferromagnetic (FM) states. As compared with LaOMAs [21], the optimized lattice constants for the corresponding oxybismuthides occur about 2-7 % larger, which is a clear manifestation of the difference between bismuth (1.82 Å) and arsenic (1.48 Å) atomic radii. The calculated stabilization energies (ΔE) of FM state relative to NM state, $\Delta E_{(FM-NM)} = E^{tot}_{FM} - E^{tot}_{NM}$, magnetic moments on 3$d$ atoms are shown in Fig.1.

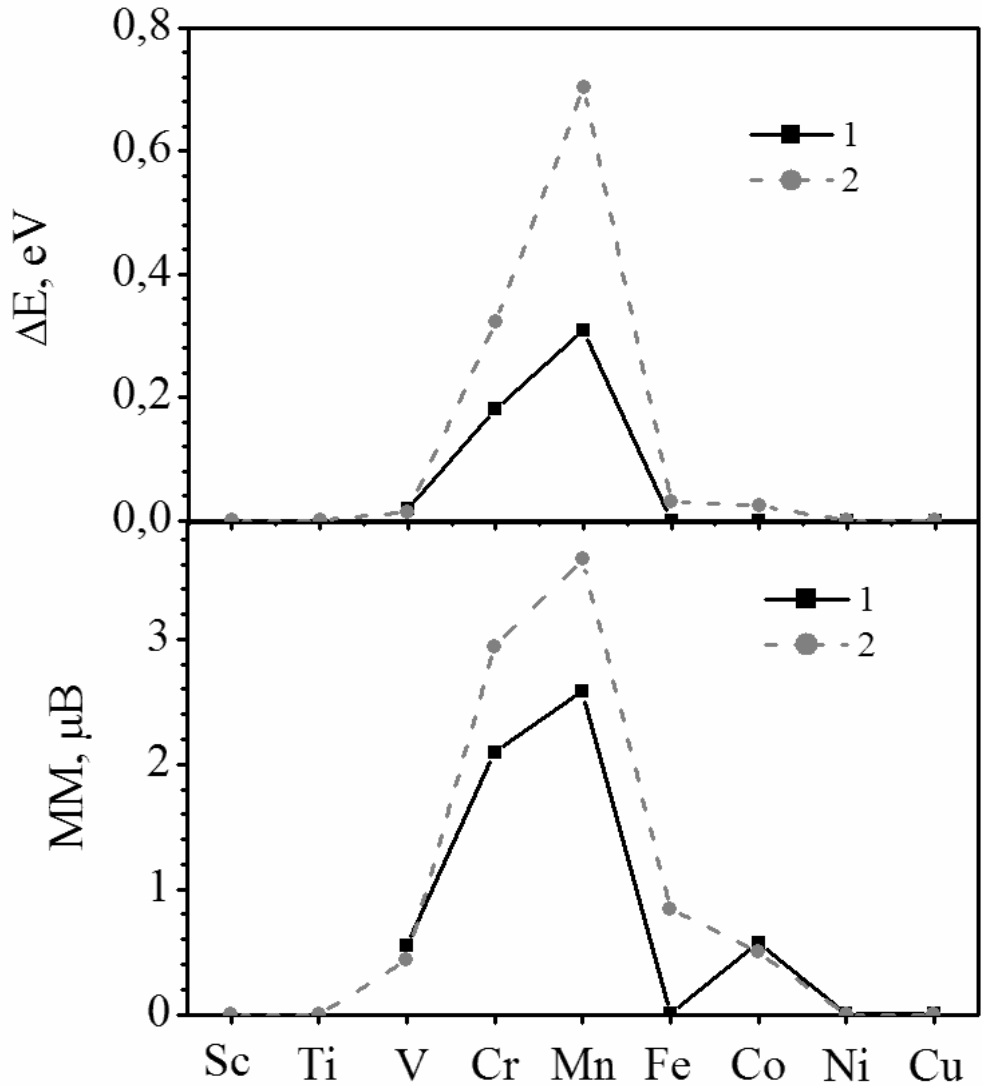

**Figure 1.** The calculated magnetic moments for 3$d$ atoms and stabilization energies for ferromagnetic states in comparison with nonmagnetic states in quaternary oxybismuthides LaOMBi, where M = Sc, Ti, V,…Ni, Cu (dotted lines) in comparison with quaternary oxyarsenides LaOMAs [20] (solid lines).

Two different groups of oxybismuthides can be clearly identified. The non-magnetic phases with M = Sc, Ti, Ni and Cu, and all others where magnetic ground state is stable. A monotonous increase of magnetic moments in magnetic LaOMBi can be seen when moving from V to Mn. This change is related to predominant filling of the spin-up states, while majority of spin-down states remain partially empty. The largest moment is observed for LaOMnBi. At the end of the 3$d$ row (Fe and Co) the filling of the 3$d$ spin-down bands begins to increase, and this leads to a monotonous decrease in the moment values so that spin splitting completely disappears in LaONiBi. Thus, magnetic (and electronic, see [24]) properties in oxybismuthides (LaO$M$Bi) and oxyarsenides (LaO$M$As [21]) seem quite similar to each other in many respects. At the same time replacement of arsenic for bismuth results in a larger cell volume [13] that, in turn, leads to some changes for 3$d$ band splitting and in increase of atomic magnetic moments for LaO$M$Bi. Most significant, qualitative difference is seen for the case of iron when LaOFeAs does not bear a magnetic moment while LaOFeBi seems to have appreciable magnetic moment on iron.

Thus, while in oxyarsenides family iron containing LaOFeAs occurs at the border of magnetic instability, the similar state for oxybismuthides ought to develop in nickel containing LaONiBi. Therefore, a preliminary conclusion can be drawn that oxybismuthide LaONiBi may possibly be considered as favorable matrix for a superconducting material.

Motivated by the calculated results, we have attempted to synthesize LaO$_{1-\delta}$NiBi. The synthetic procedure included smelting of high purity La and Bi metals to LaBi metal precursor at the starting stage. Then carbonyl nickel metal powder and nickel oxide were added to the precursor in proportions nominally corresponding to $\delta = 0.2$, thoroughly mixed together in a dry box, pressed in pellets, wrapped in tantalum foil, placed in silica ampoule, evacuated and sealed. The ampoule was fired at 800°C for 10 hours. The procedure was repeated to obtain specimens with X-ray diffraction pattern shown in Fig.2.

A small amount of impurity phases (about 5%) are present in the samples. The diffraction peaks were indexed with P4/nmm space group. The tetragonal elementary unit parameters were found equal to $a = 4.073$ and $c = 9.301$ Å.

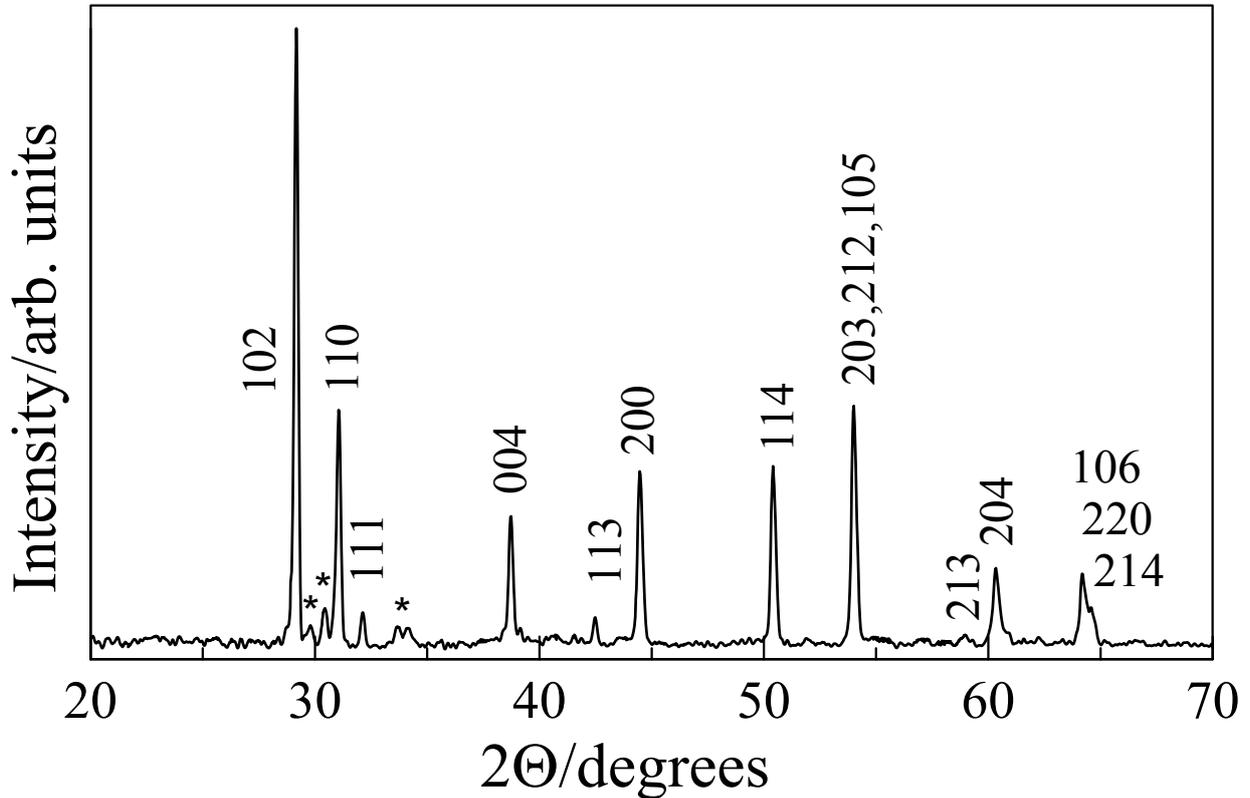

**Figure 2**. Powder X-ray diffraction pattern for as-synthesized LaO$_{0.8}$NiBi. Impurities are shown with asterisks.

Resistivity $\rho$ and Hall coefficient $R_H$ were measured by standard four-probe technique within the temperature range 1.5 – 300 K and in magnetic field up to 13.6 T. The specimens show a narrow superconducting transition at $T_c = 4.25$ K measured by both resistivity and ac-susceptibility methods with almost 100% shielding at T ≤ 3 K. The upper critical field H$_{c2}$, defined at 90% of normal state resistivity, shows approximately linear $T$-dependence with the slope dH$_{c2}$/d$T \approx 1$ T/K, Fig.3, which is similar to that observed in nickel oxyarsenide LaONiAs [25]. The changes in $\rho$ and R$_H$ with Temperature reveal complex behavior, which is typical for two-band metals. R$_H(T)$ is positive and shows a strong decrease with the temperature increase. Similar behavior was observed in Sr-doped LaOFeAs [8].

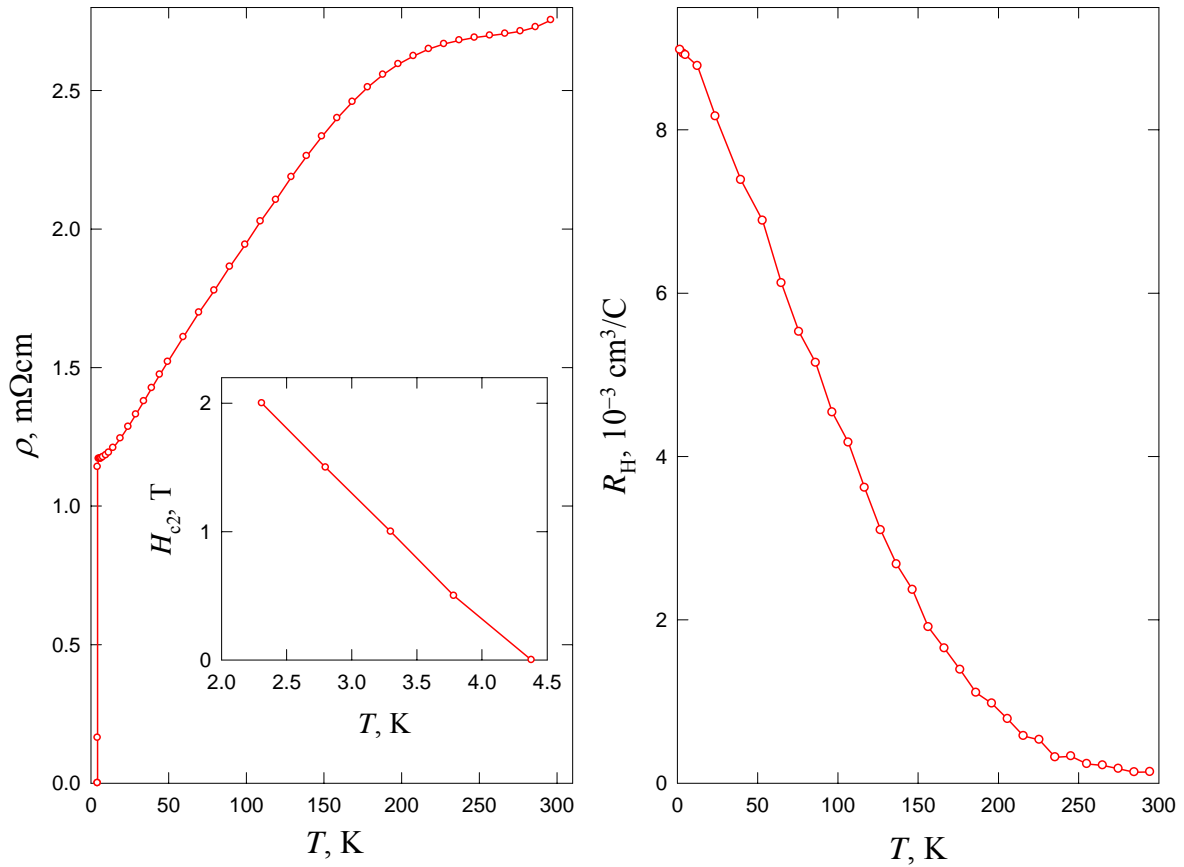

**Figure 3**. Temperature dependencies of resistivity $\rho$ (left panel), Hall coefficient $R_H$ (right panel) and upper critical field $H_{c2}$ (insert).

In summary, the oxide compound $LaO_{0.8}NiBi$ was first synthesized via solid state route. The resemblance in structure and electronic properties, the temperature changes in conductivity and in upper critical field similar to those in oxyarsenides are all indicative of the identical mechanism that underlies transition to superconducting state in oxyarsenides and oxybismuthides.

This work was carried out with partial financial support from the Programs for Basic Research of Presidium of RAS "Quantum Macrophysics" and "Effect of Atomic-Crystal and Electron Structure on Condensed Matter Properties", RFBR grant No.07.02.00020 and State Contract No. 02.518.11.7026.